\documentstyle[epsfig]{article}

\catcode`\@=11 
\def\@aabuffer{}
\def\author #1{\expandafter\def\expandafter\@aabuffer\expandafter
{\@aabuffer \small\rm      #1\relax \par}}
\def\address#1{\expandafter\def\expandafter\@aabuffer\expandafter
{\@aabuffer \small\it #1\relax \par\vspace{1em}}}

\def\maketitle{
\begin{center}
   {\bf \@title \par}       
   \vskip 2em                      
   \@aabuffer\relax
\end{center} \par
\gdef\@aabuffer{}
}

\def\abstracts#1{
\begin{center}
{\begin{minipage}{4.2truein}
                 \footnotesize
                 \parindent=0pt #1\par
                 \end{minipage}}\end{center}
                 \vskip 2em \par}
\def\@citex[#1]#2{\if@filesw\immediate\write\@auxout
	{\string\citation{#2}}\fi
\def\@citea{}\@cite{\@for\@citeb:=#2\do
	{\@citea\def\@citea{,}\@ifundefined
	{b@\@citeb}{{\bf ?}\@warning
	{Citation `\@citeb' on page \thepage \space undefined}}
	{\csname b@\@citeb\endcsname}}}{#1}}

\newif\if@cghi
\def\cite{\@cghitrue\@ifnextchar [{\@tempswatrue
	\@citex}{\@tempswafalse\@citex[]}}
\def\citelow{\@cghifalse\@ifnextchar [{\@tempswatrue
	\@citex}{\@tempswafalse\@citex[]}}
\def\@cite#1#2{{$\!^{#1}$\if@tempswa\typeout
	{IJCGA warning: optional citation argument 
	ignored: `#2'} \fi}}

\renewenvironment{thebibliography}[1]
	{\begin{list}{\arabic{enumi}.}
	{\usecounter{enumi}\setlength{\parsep}{0pt}
	 \setlength{\itemsep}{0pt} 
         \settowidth
	{\labelwidth}{#1.}\sloppy}}{\end{list}}

\long\def\@makecaption#1#2{
   \vskip 10pt 
   \setbox\@tempboxa\hbox{\footnotesize #1: #2}
 \ifdim \wd\@tempboxa >\hsize \footnotesize #1: #2\par \else \hbox
 to\hsize{\hfil\box\@tempboxa\hfil}  
   \fi}

\catcode`\@=12

\def\Journal#1#2#3#4{{#1} {\bf #2}, #3 (#4)}

\def\NPB{{\em Nucl. Phys.} B}
\def\PLB{{\em Phys. Lett.}  B}
\def\PRL{\em Phys. Rev. Lett.}
\def\PRD{{\em Phys. Rev.} D}
\def\ZPC{{\em Z. Phys.} C}
\def\EPJC{{\em Europ. Physical Journal} C}
\def\APL{{\em Acta Phys. Pol.} B}

\def\be{\begin{equation}}
\def\ee{\end{equation}}
\def\bea{\begin{eqnarray}}
\def\eea{\end{eqnarray}}

\begin{document}

\rightline{BI-TP 2000/16}
\rightline{MPI-PhT 2000-23}\bigskip
                        
\title{OFF-DIAGONAL GENERALIZED VECTOR-DOMINANCE
AND COLOUR-DIPOLES IN LOW-x DIS\footnote{Contribution to DIS2000, Liverpool,
April 2000}}

\author{D. SCHILDKNECHT}

\address{
Fakult\"at f\"ur Physik, Universit\"at Bielefeld,\footnote{Permanent 
address}\\             
Universit\"atsstra{\ss}e 25, 33615 Bielefeld\\
and
MPI f\"ur Physik, M\"unchen\\
E-mail: Dieter.Schildknecht@physik.uni-bielefeld.de}

\maketitle

\abstracts{
We briefly summarize the equivalence of
off-diagonal generalized vector dominance and the colour-dipole
approach to deep-inelastic scattering (DIS) in the diffraction
region of values of $x \simeq Q^2/W^2 << 1$.}

I am happy to be back at Liverpool. Actually, it is the first time
after my participation at the 1969 International Symposium on Electron
and Photon Interactions. At the time of the 1969 Symposium, the
interpretation of electromagnetic interactions of the hadrons in terms
of vector-meson dominance \cite{Sakurai} was at its height. It was,
e.g., theoretically conjectured \cite{Stodo} and experimentally
confirmed \cite{Sakurai} that the total cross section for photoproduction
on nucleons was quantitatively in good approximation related to, and
explained in terms of, (diffractive) vector-meson ($\rho^0, \omega, \phi$)
scattering and ($\rho^0, \omega, \phi$) photoproduction.

\noindent
\begin{figure}[ht]
\vskip 3mm \hskip 1cm \epsfig{figure=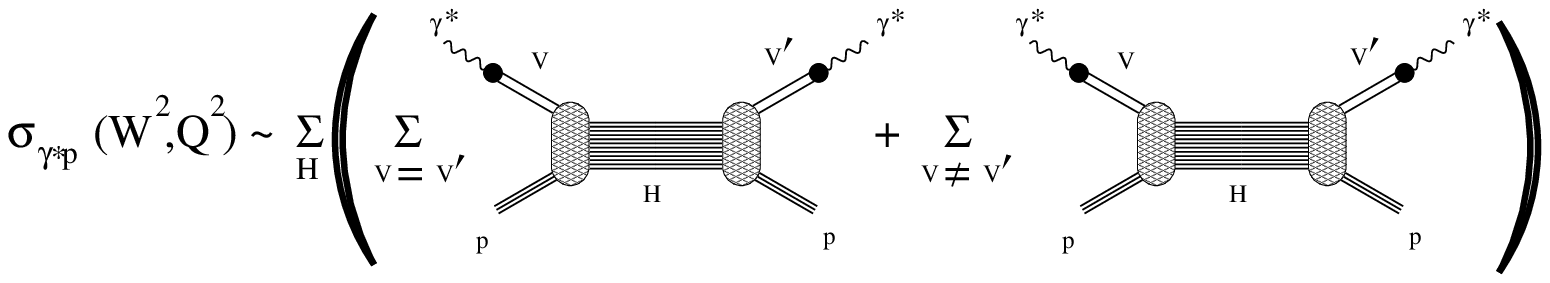,height=15mm}
\caption{The imaginary part of the forward Compton amplitude in GVD
\label{Fig1}}
\end{figure}

The 1969 Symposium also saw the rise of the parton model \cite{Gilman},
and the failure of pure $(\rho^0, \omega, \phi)$ dominance to
explain DIS, or, equivalently, the photoproduction cross section, as soon
as the photon acquired a spacelike four momentum, $Q^2 >> m^2_\rho$.
Indeed, it was experimentally established that $\sigma_{\gamma x p}
(W^2, Q^2)$ behaved as $1/Q^2$ rather than fulfilling the $\rho^0$-dominance
prediction of $m^4_\rho/Q^4$. Conjecturing the transition of the photon
to more massive vector states, experimentally unknown at that time, and their
subsequent diffractive interaction with the nucleon, lead to generalized
vector dominance (GVD) \cite{Sak-Schi} as a theory for DIS in the low
$x \simeq Q^2/W^2 << 1$ region, see Fig. \ref{Fig1}.

\noindent
\begin{figure}[ht]
\parbox[l]{5cm}{
 \epsfig{file=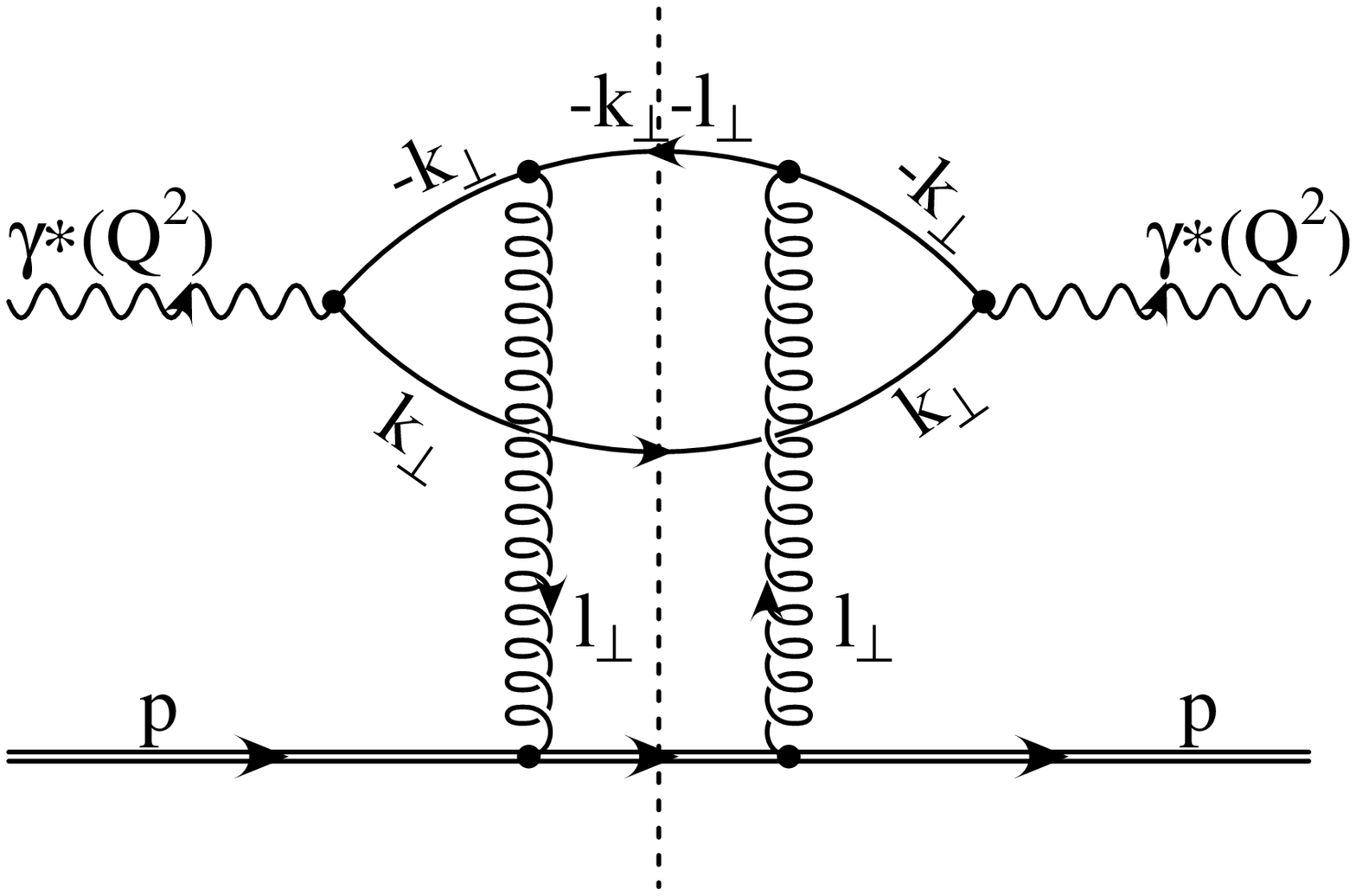,width=5.0cm}
 \vskip 1.5 cm
 \epsfig{file=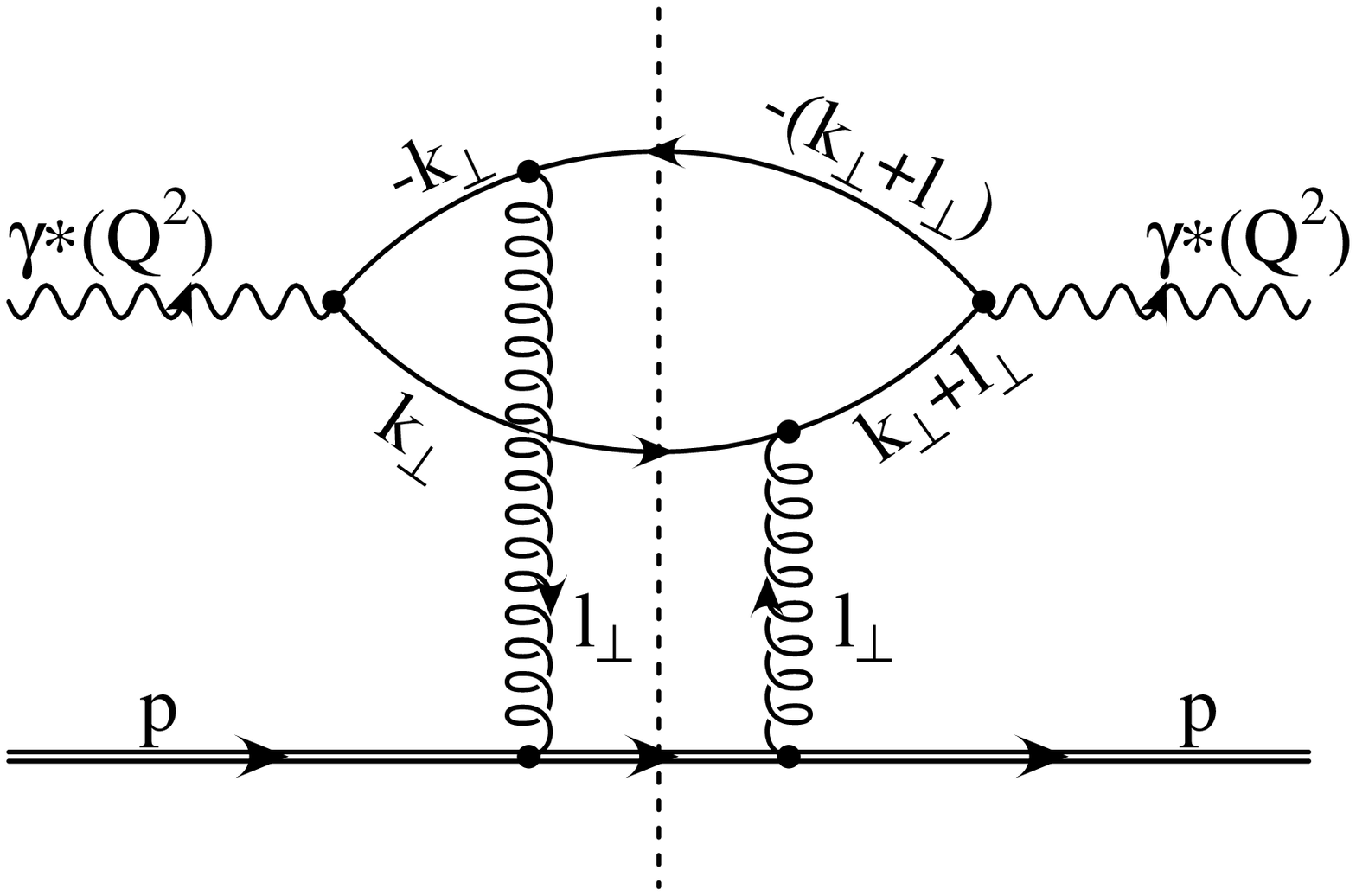,width=5.0cm}
\caption{The two-gluon exchange
\label{Fig2}}
}
\parbox[r]{6cm}{
\epsfig{file=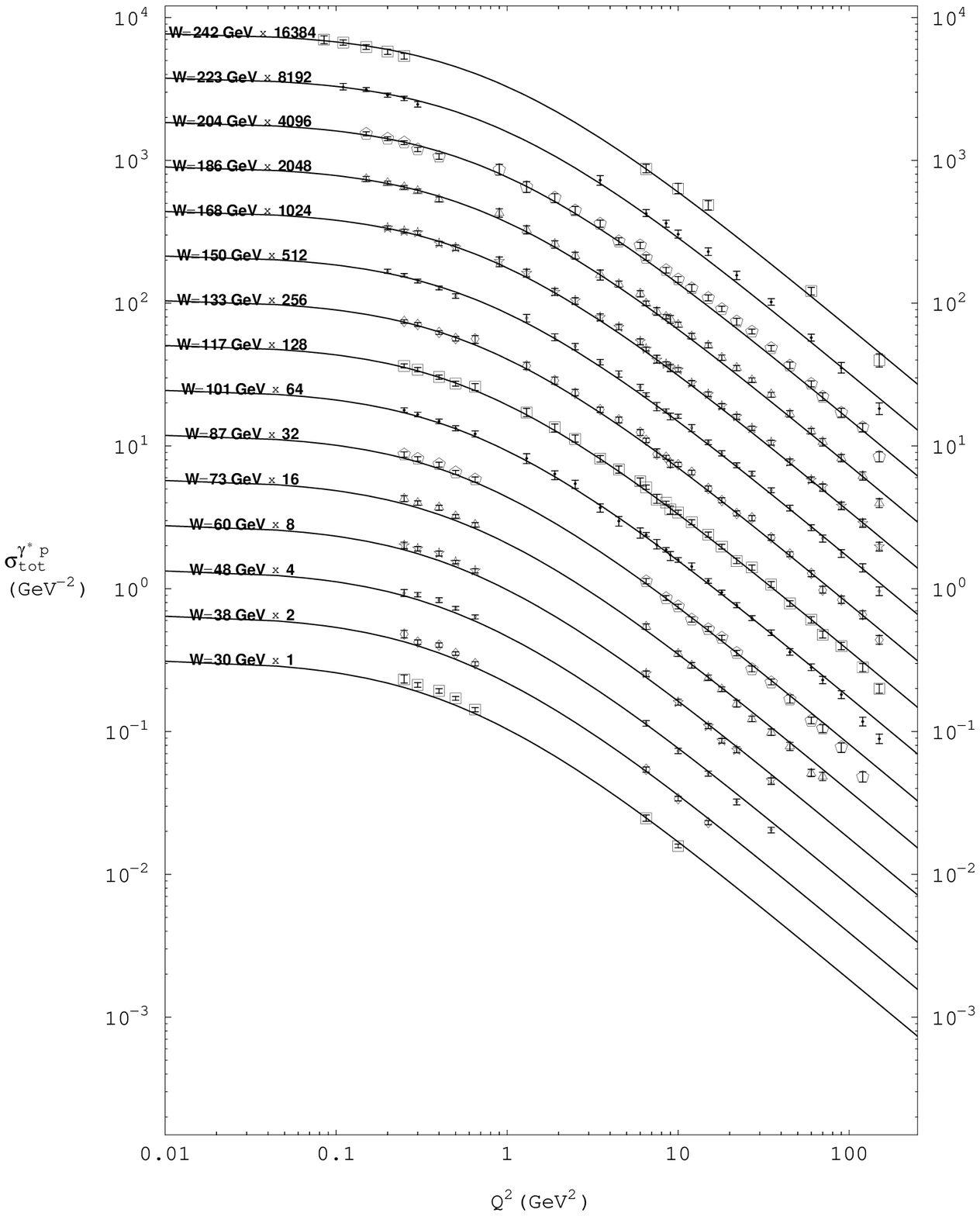,width=6.cm}
\caption{ Colour-dipole or, equivalently, GVD results \protect\cite{Tent} for
$\sigma_{\gamma^*p}~(W,Q^2)$ compared with ZEUS data \protect\cite{ZEUS}.
\label{Fig3}}
}                             
\end{figure}

It only has been in recent years, with the advent of HERA, that the
conceptual basis of the GVD approach was experimentally confirmed by the
observation of diffractive production of high-mass $(q \bar q)$ states
in DIS at small values of $x << 1$ and all $Q^2$.
As depicted in Fig. \ref{Fig1}, the forward Compton amplitude contains diagonal
as well as off-diagonal transitions with respect to (the masses of) the
ingoing and outgoing $(q \bar q)$ vector states. Inconsistencies
(sometimes called \cite{Bjorken} the ``Gribov paradox''),
occuring when for simplicity off-diagonal transitions are ignored,
twentyfive years
ago, lead us to seriously consider \cite{Fraas} off-diagonal transitions
with destructive interference between diagonal and off-diagonal
contributions. The structure of the forward Compton amplitude thus arrived
at \cite{Fraas}, found an a posteriori justification \cite{Cvetic}
in perturbative QCD by the
generic structure of two-gluon exchange \cite{Gunion},
compare Fig. \ref{Fig2}.
It turned out that off-diagonal GVD is largely equivalent
to\cite {Cvetic,Frank} what has become known as the colour-dipole
approach \cite{Nikolaev}. Our results from ref. \cite{Cvetic} were recently
generalized \cite{Tent} to include the $W$ dependence, compare Fig. \ref{Fig3}.
While our results are formulated in momentum space, closely related
work \cite{Golec,Forshaw} is based on transverse position space, compare
the talks \cite{Shaw} by Krzysztof Golec-Biernat and
Graham Shaw at this meeting.

In his talk \cite{Levy} on ``The Legacy of HERA'' at this meeting,
Aharon Levy mentioned that he was
``shocked'' when he first saw diffractive high-mass events appearing at HERA
at small $x$ and any $Q^2$. Some may have been shocked, others not.
As for myself, I allow myself to say that I was not shocked at all, as I 
saw our theoretical conjectures \cite{Sak-Schi,Fraas} confirmed.

\section*{Acknowledgments}
It is a pleasure to thank Gorazd Cvetic, Arif Shoshi and Mikhail Tentioukov
for a pleasant collaboration and my friends and colleagues at the MPI in 
Munich for warm hospitality.

\end{document}